# Statistically Characterizing the Electrical Parameters of the Grid Transformers and Transmission Lines [*]

Mir Hadi Athari
Department of Electrical and Computer Engineering
Virginia Commonwealth University
Richmond, VA, USA
Email: atharih@vcu.edu

Zhifang Wang[†]
Department of Electrical and Computer Engineering
Virginia Commonwealth University
Richmond, VA, USA
Email: zfwang@vcu.edu

*Abstract*—**This paper presents a set of validation metrics for transmission network parameters that is applicable in both creation of synthetic power system test cases and validation of existing models. Using actual data from two real world power grids, statistical analyses are performed to extract some useful statistics on transformers and transmission lines electrical parameters including per unit reactance, MVA rating, and their X/R ratio. It is found that conversion of per unit reactance calculated on system common base to transformer own power base will significantly stabilize its range and remove the correlation between per unit X and MVA rating. This is fairly consistent for transformers with different voltage levels and sizes and can be utilized as a strong validation metric for synthetic models. It is found that transmission lines exhibit different statistical properties than transformers with different distribution and range for the parameters. In addition, statistical analysis shows that the empirical PDF of transmission network electrical parameters can be approximated with mathematical distribution functions which would help appropriately characterize them in synthetic power networks. Kullback-Leibler divergence is used as a measure of goodness of fit for approximated distributions.**

## I. Introduction

Synthetic power networks are emerging as a potential solution for the lack of test cases for performance evaluation in power system research and development. Generally, access to real data in critical infrastructure like power networks is limited due to confidentiality requirements. Utility companies and regulatory agencies don't share such data and strictly limit access to actual power systems data for public and researchers due to their sensitivity. On the other hand, it is important that new concepts and algorithms developed by researchers be evaluated in relatively large and complex networks with the same characteristics as actual grids so that they can be reproducible by peers. For example, authors in [1]–[3] have developed a new storage management and energy management algorithms which enable a bidirectional power flow from microgrids to power networks that need evaluation with realistic grid topology. Since Synthetic power networks are entirely fictitious but with the same characteristics as realistic networks, they can be freely published to the public to facilitate advancement of new technologies in power systems.

Development of efficient synthetic power system models requires that their size, complexity, and electrical and topological characteristics match those of real power grids. Power networks are complex infrastructures with various components. In addition to topological characteristics of power networks, they include several components with different electrical characteristics such as different types of transformers, switched shunt reactive power compensation, remote tap changing bus voltage regulation, etc. Development of synthetic power networks with the same complexity that can simulate the exact behavior of actual grids needs a comprehensive study of different components from both electrical and topological perspectives. Also, increasing level of renewable generation in power systems has introduced an unprecedented level of uncertainty into grids [4]. In the literature, many studies are dedicated for characterizing actual power networks mainly from topological perspectives such as ring-structured power grid developed in [5] and tree structured power grid model to address the power system robustness [6], [7]. Small world approach described in [8] served as a reference for the works of [9]–[11] to develop an approach for generating truly synthetic transmission line topologies. A random topology power network model, called *RT-nestedSmallWorld*, is proposed in [10] based on comprehensive studies on the electrical topology of some real world power grids. The impacts of different bus type assignments in synthetic power networks on grid vulnerability to cascading failures are investigated in [12].

In [13] the authors presented a substation placement method and transmission lines assignment from real energy and population data based on methodology introduced in [14], [15]. The proposed methodology employs a clustering technique to ensure that synthetic substations meet realistic proportions of load and generation. However, the authors will continue to augment test cases by adding additional complexities such as transmission network electrical parameters assignment. In another study, the authors performed a statistical analysis on transmission line capacity regarding both topology and electrical parameters. However, all these studies focus mainly on topology-related parameters of transmission lines and ignore electrical parameters such as impedance of transmission lines and transformers.

Review of the literature on synthetic grid modeling reveals that there is a need for statistical studies to characterize electrical parameters of transmission network to be used in synthetic grid models. In this paper, we mainly focus on the statistical analysis of transformers and transmission lines electrical parameters such as per unit impedance, nominal capacity and X/R ratio. The goal of this paper is to a) provide a well-defined "rules" for transmission network parameters as potential validation metrics for existing synthetic grid models and b) to provide guidelines on how to accurately configure them in synthetic models. A very large sample of actual operating transformers and transmission lines from two real world power systems is used to extract the statistical characteristics of their parameters.

The rest of the paper is organized as follows. Section II presents the statistical analysis on transformers electrical parameters. Section III discusses the statistics of transmission lines parameters and finally some concluding remarks and future work direction will be presented in section IV.

## II. GRID TRANSFORMERS

Generally, in power systems branches are referred to transmission lines or transformers between two buses in the network. Also, in some cases shunts are considered in the branch category. In this paper we first perform some statistical analysis on transformers electrical parameters extracted from two real world power systems. Next, transmission lines from the same networks will be studied to extract some statistics for their critical parameters.

### A. Per unit impedance using the system MVA base or transformer's power rating?

In power system analysis the use of per unit system to express the system quantities as fractions of a defined base unit quantity is common. This is important especially for transformers as the voltage level is different for their terminals and per unit system simplifies transformer calculations. Another advantage for this expression is that similar types of apparatus like transformers will have the impedances lying within a narrow numerical range when expressed as a per-unit fraction of the equipment rating, even if the unit size varies widely. However, per unit impedances of power grid components are usually converted to new values using a common system-wide base for application in power system analysis like power flow or economic power flow calculations. This conversion depends on reference voltage base for different zones in the system and a predefined unique power base for the entire system according to the following simple equation:

$$Z_{PU}^{New} = Z_{PU}^{Given} \times \left(\frac{V_{Base}^{Given}}{V_{Base}^{New}}\right)^2 \times \left(\frac{S_{Base}^{New}}{S_{Base}^{Given}}\right) \quad (1)$$

where $Z_{PU}^{Given}$, $V_{Base}^{Given}$, $S_{Base}^{New}$ are given per unit impedance, voltage base, and power base for each apparatus and $Z_{PU}^{New}$ is the new per unit impedance calculated using $V_{Base}^{New}$ and $S_{Base}^{New}$. Usually, the voltage base values are selected the same as the nominal voltage of transformer terminals for each zone to simplify the calculations. Therefore, the conversion formula for per unit impedance can be expressed as

$$Z_{PU}^{New} = Z_{PU}^{Given} \times \left(\frac{S_{Base}^{New}}{S_{Base}^{Given}}\right) \quad (2)$$

In the power grids the use of different voltage levels is a common practice to decrease the power loss through transmission lines. Thus there are transformers with different turn ratios to couple the areas with different voltage levels. In this study, the transformers are grouped into different categories based on their high voltage terminals. This is because as the nominal voltage level increases the transformer size gets larger, so studying them in groups based on voltage level seems reasonable for extracting validation metrics. The purpose of statistical experiments in this study is to identify several validation metrics for transformers parameters including their impedances to help validate synthetic power networks. This would be even more helpful if the range for different parameters can be specified for typical power system components. The first experiment tries to find the relationship between MVA rating of transformers and their per unit impedances. These analyses are performed on both per unit values in system base, and converted values to transformers own MVA ratings. The original power system data used in this study offer transformer impedance in per unit calculated based on the common base for the system. Fig. 1 shows the scatter plot of transformers per unit reactance (X) and MVA rating for the original and converted per unit reactance of transformers. Note that although transformers with high voltage terminal of 115 kV are selected for this comparison, the results are fairly consistent for other voltage levels as shown in fig. 2.

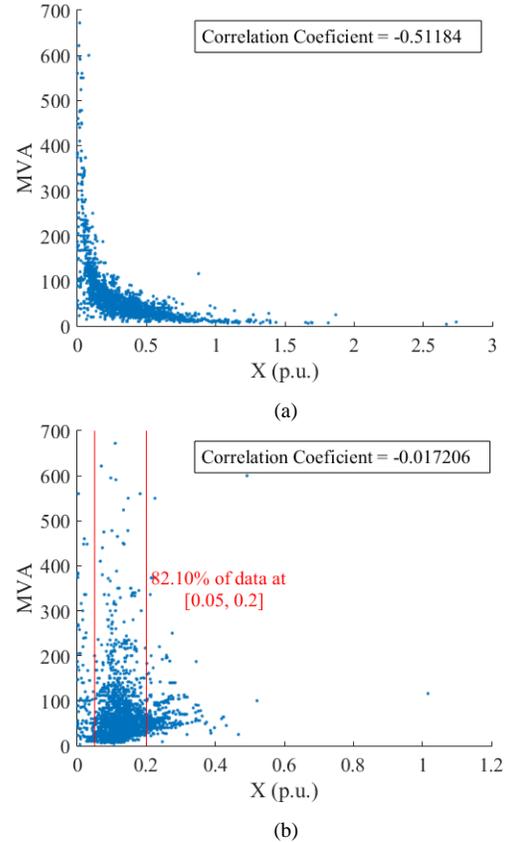

Figure 1. Scatter plot of per unit reactance versus MVA rating of transformer for a) system common base and b) converted to transformer own MVA rating.

The scatter plot for per unit reactance on system common base shows a descending trend as the size of transformer increases which means there is relatively large correlation coefficient between the two as shown in fig. 1 (a). In this case, the per unit reactance values span from nearly 0 to 2.75 p.u which is relatively large range for this parameter. However, when we consider the same scatter plot for converted per unit reactance to transformer own MVA rating, this range narrows down to [0, 0.5] p.u putting at least 80% of them within even a narrower range of [0.05, 0.2] p.u. In addition, almost zero correlation coefficient means that this range is independent of transformer size and voltage level.

The same scatter plots for converted values of per unit reactance versus MVA rating of transformers for other voltage levels are depicted in fig.2. It is found that per unit reactance of transformers in power systems regardless of their size lie within a narrow range when calculated on their own power base and statistics reflect what is known from engineering practice. This can be a potential validation metric for synthetic power networks transformers along with other statistical measures such as their probability distribution.

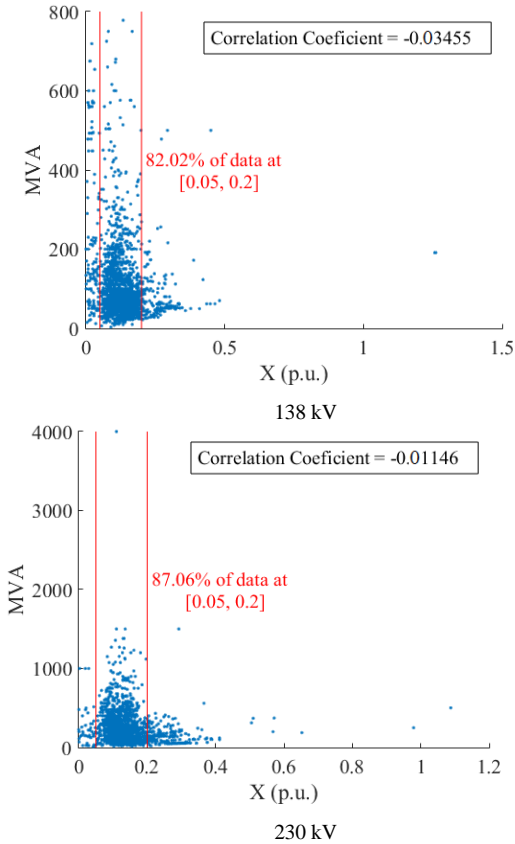

Figure 2. Scatter plot of per unit reactance versus MVA rating of transformer for 138 and 230 kV transformers.

### B. Transformer parameter distribution

Transformer parameters statistics are derived using over 30000 actual power transformers. The database includes different types of transformers such as fixed step down and step up transformers, three winding transformers, On-load Tap Changer (OLTC) transformers, and autotransformers. A negative impedance often occurs in the star modeling of a three winding transformer due to how the leakage reactance is measured/modeled [16]. Also, Network equivalencing methods can create negative impedances which can affect the statistics of transmission network parameters. To avoid such scenario, data are filtered by $R > 0, X > 0$ to exclude abnormal transformer parameters from samples. Also, due to lack of detailed information on some transformers, their MVA ratings are reported with either very large or zero values. These transformers too are excluded from samples to have accurate statistics.

The probability distribution of transformer parameters is another measure that can be used along with parameter range as validation metric in synthetic power networks. The probability distribution of a random variable, say transformer per unit reactance, is a function that describes how likely we can obtain the different possible values of the random variables. Using the database of real transformer data, we can get the empirical cumulative density function (CDF) of each parameter that can give us the empirical probability density function (PDF). Next, to provide a more systematic approach for generating synthetic models, we try to fit approximated distribution functions to empirical PDFs. The goodness of this fit can be measured with Kullback-Leibler divergence.

*1) Kullback-Leibler Divergence*

In probability theory and information theory, the Kullback–Leibler (KL) divergence, also called discrimination information, is a measure of the difference between two probability distributions P and Q. It is not symmetric in P and Q. In applications, P typically represents the "true" distribution of data, observations, or a precisely calculated theoretical distribution, while Q typically accounts for a theory, model, description, or approximation of P [17]. Specifically, the KL divergence from Q to P, denoted $D_{KL}(P \parallel Q)$, is the amount of information lost when Q is used to approximate P. For discrete probability distributions P and Q, the KL divergence from Q to P is defined to be [18]

$$D_{KL}(P \parallel Q) = \sum_i P(i) log \frac{P(i)}{Q(i)} \qquad (3)$$

In words, it is the expectation of the logarithmic difference between the probabilities P and Q, where the expectation is taken using the probabilities P. Therefore, smaller values for the divergence represents more accurate fit for the empirical PDF of transformer parameters.

*2) Transformer per unit reactance distribution*

Three different voltage levels, 115, 138, and 230 kV are selected to report in this study. Transformers are grouped based on their high voltage terminal and categorized into three voltage levels. For transformer per unit reactance, the converted per unit values to transformer power base is used to identify the distribution of per unit reactance. Fig. 3 shows the empirical PDF of transformer per unit reactance for different voltage levels. As found earlier in this paper, converted values of per unit reactance lie within a narrow range. According to KL divergence measure, it is found that the distribution of per unit reactance can be approximated with t Location-Scale (TLS) distribution with three parameters as shown in the following distribution function:

$$f(x|\mu,\sigma,\nu) = \frac{\Gamma\left(\frac{\nu+1}{2}\right)}{\sigma\sqrt{\nu\pi}\Gamma\left(\frac{\nu}{2}\right)}\left[\frac{\nu + \left(\frac{x-\mu}{\sigma}\right)^2}{\nu}\right]^{-\left(\frac{\nu+1}{2}\right)} \quad (4)$$

where $\Gamma$ is the gamma function, $\mu$ is the location parameter, $\sigma$ is the scale parameter, and $\nu$ is the shape parameter. The mean of the TLS distribution is $\mu$ and is only defined for $\nu > 1$ and the variance is $var = \sigma^2 \frac{\nu}{\nu-2}$ and is only defined for $\nu > 2$. Note that if random variable $x$ has a TLS distribution with parameters $\mu$, $\sigma$, and $\nu$, then $\frac{x-\mu}{\sigma}$ has a Student's t-distribution with $\nu$ degrees of freedom. In probability and statistics, Student's t-distribution (or simply the t-distribution) is any member of a family of continuous probability distributions that arises when estimating the mean of a normally distributed population in situations where the sample size is small.

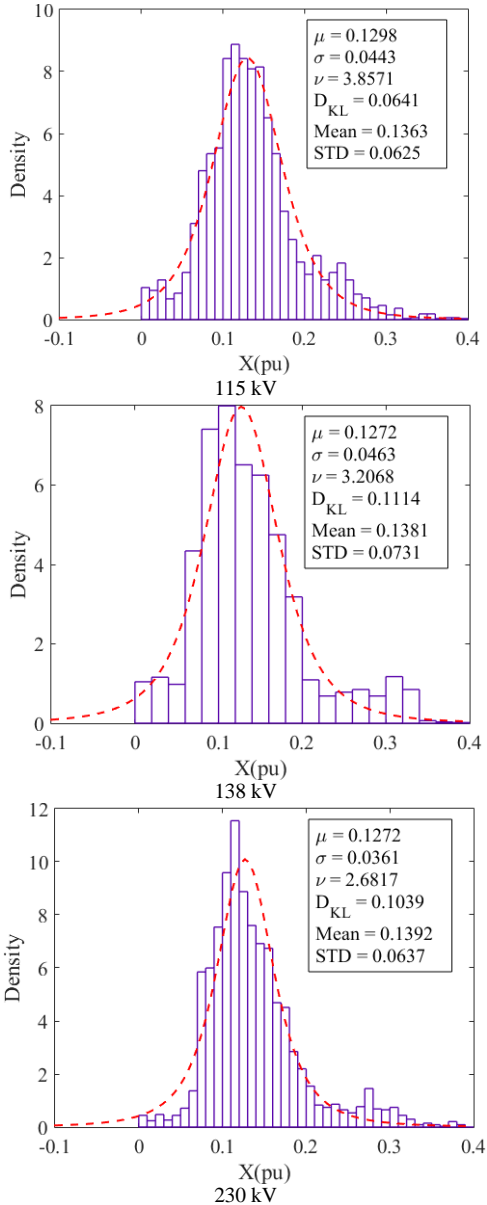

Figure 3. Empirircal PDF and TLS-fit of per unit reactance for 115, 138, and 230 kV transformers.

Table I shows the median, mean, minimum and maximum range, and the percentage of per unit reactance lying within [0.05, 0.2] p.u range for the two real world power grids.

TABLE I. PER UNIT REACTANCE STATISTICS FOR 115, 138, AND 230 KV TRANSFORMERS

| Voltage Levels (kV) | Transformer Per unit reactance | | | |
|---|---|---|---|---|
| | Median | Mean | Range | % at [0.05, 0.2] |
| 115 | 0.1291 | 0.1363 | [3.92e-4, 1.0162] | 81.88 |
| 138 | 0.1246 | 0.1381 | [1.00e-4, 1.26] | 82.01 |
| 230 | 0.1260 | 0.1392 | [2.47e-4, 1.08] | 87.33 |

*3) Transformer Capacity Distribution*

Another key parameter of a transformer is its capacity or MVA rating. For the set of data from real world power grids, there are transformers with different sizes from couple MVA to +1000 MVA. Also, due to the lack of detailed information in some cases, the MVA rating of some transformers are set to a very large or small values. To exclude such cases, in addition to identifying the full range of transformer MVA rating, an 80% range centered at the median is defined to get rid of "extreme values" on both upper and lower bounds. This will give us a more useful range where most transformers fall in. Table II shows the median, mean, minimum and maximum range, and 80% range for transformers MVA ratings.

TABLE II. MVA RATING STATISTICS FOR 115, 138, AND 230 KV TRANSFORMERS

| Voltage Levels (kV) | Transformer MVA rating | | | |
|---|---|---|---|---|
| | Median | Mean | Range | 80% range |
| 115 | 53 | 71.30 | [3, 384] | [22, 140] |
| 138 | 83 | 117.24 | [3.3, 616] | [39, 239] |
| 230 | 203 | 246.61 | [10, 1380] | [62.5, 470] |

Fig. 4 depicts the empirical PDF of transformers MVA rating and the approximated fit distribution for 115 kV transformers. Note that the results for 138 kV and 230 kV transformers will be presented later in a table. According to the KL divergence, transformers capacity is approximated with Generalized Extreme Value (GEV) distribution with the minimum $D_{KL}$ value where its CDF is represented by (5)

$$F(x|\zeta,\mu,\sigma) = exp\left(-\left(1 + \zeta\frac{(x-\mu)}{\sigma}\right)^{\frac{-1}{\zeta}}\right) \quad (5)$$

where $\mu$ is location parameter, $\sigma$ is scale parameter, and $\zeta \neq 0$ is shape parameter. Using this mathematical distribution, one can generate reasonable values for transformer capacities in a given synthetic grid model.

*4) Transformer X/R distribution*

The third important parameter of transformers is the ratio of their per unit reactance to per unit resistance. Using such ratio, one can estimate the value of per unit resistance given the range and distribution of per unit reactance of the transformer. These two parameters form the real and imaginary parts of transformer impedance that is necessary for power flow analysis in synthetic power networks.

Table III shows the median, mean, minimum and maximum range, and 80% range for transformers MVA ratings. The 80%

range is determined using the same approach as used in MVA rating determination.

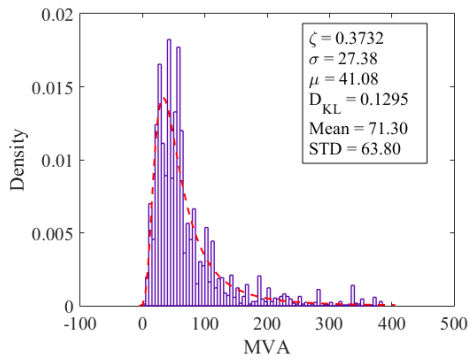

Figure 4. Empirircal PDF and GEV-fit of MVA rating for 115 kV transformers.

TABLE III. X/R RATIO STATISTICS FOR 115, 138, AND 230 KV TRANSFORMERS

| Voltage Levels (kV) | Transformer MVA rating | | | |
|---|---|---|---|---|
| | Median | Mean | Range | 80% range |
| 115 | 25.39 | 37.83 | [0.0577, 5.41e3] | [16.2, 47.5] |
| 138 | 29.58 | 39.73 | [0.2033, 1.92e3] | [19.1, 54] |
| 230 | 44.37 | 65.77 | [0.1786, 4.03e3] | [25, 84] |

Fig. 5 shows the empirical and approximated distribution for 115 kV transformers. Again, it is found that the GEV distribution can fit the data best according to KL divergence measure. Some very small X/R ratios come from autotransformers, and the ballpark is that if the ratio is less than 4 to 1, it is an autotransformer.

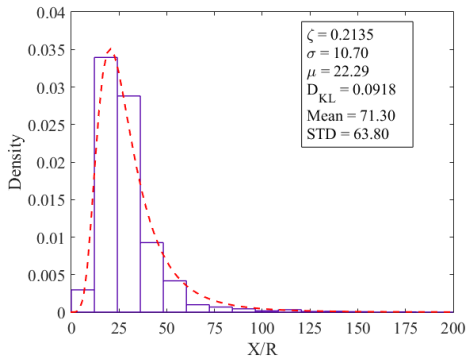

Figure 5. Empirircal PDF and GEV-fit of X/R ratio for 115 kV transformers.

Table IV presents the estimated parameters of the best fitting functions for transformer MVA rating and X/R ratio.

### III. TRANSMISSION LINES

Transmission line parameters statistics are derived using over 50000 lines from real power systems. Transmission lines are categorized based on their nominal voltage level which ranges from 0.6 to 765 kV. Here we study lines with nominal voltage levels of 115, 138, and 230 kV. We studied per unit reactance, X/R ratio, and line capacities as three critical parameters of transmission lines to provide several validation metrics and guidelines for synthetic grid modeling.

### A. Transmission line per unit reactance distribution

Fig. 6 shows the empirical PDF of transmission line per unit reactance and the approximated fit distribution for different voltage levels.

TABLE IV. THEORETICAL DISTRIBUTION FUNCTIONS ESTIMATED PARAMETERS FOR TRANSFORMERS MVA RATING AND X/R RATIO

| | Estimated Parameters | | |
|---|---|---|---|
| | 115 kV | 138 kV | 230 kV |
| MVA rating statistics | $D_{KL}$= 0.1295 | $D_{KL}$= 0.0990 | $D_{KL}$= 0.1148 |
| | $\mu$= 41.08 | $\mu$= 66.82 | $\mu$= 154.79 |
| | $\sigma$= 27.38 | $\sigma$= 42.31 | $\sigma$= 105.61 |
| | $\zeta$= 0.3732 | $\zeta$= 0.4166 | $\zeta$= 0.2433 |
| X/R ratio statistics | $D_{KL}$= 0.0918 | $D_{KL}$= 0.0949 | $D_{KL}$= 0.0984 |
| | $\mu$= 22.29 | $\mu$= 25.88 | $\mu$= 37.79 |
| | $\sigma$= 10.70 | $\sigma$= 12.34 | $\sigma$= 19.67 |
| | $\zeta$= 0.2135 | $\zeta$= 0.2167 | $\zeta$= 0.2594 |

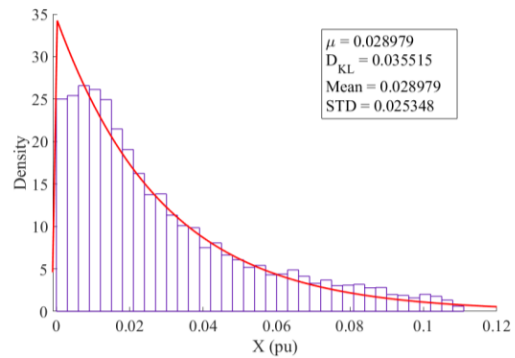

115 kV

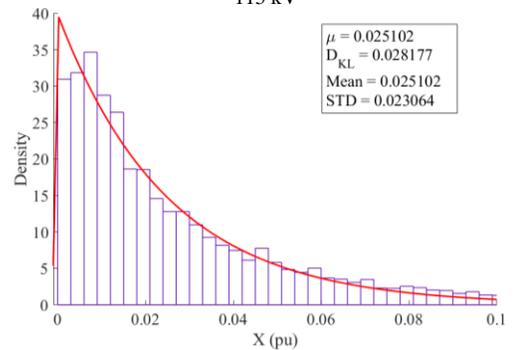

138 kV

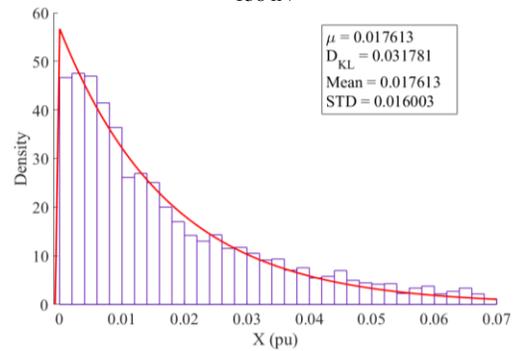

230 kV

Figure 6. Empirircal PDF and Exponential-fit of per unit reactance for 115, 138, and 230 kV transmission lines.

It is found that for all three voltage levels, per unit reactance is mostly less than 0.02 p.u. and the density drops exponentially as reactance increases. According to the KL divergence, transmission line reactance is approximated with Exponential distribution with the minimum $D_{KL}$ value where its PDF is represented by (6)

$$f(x|\mu) = \frac{1}{\mu} e^{\frac{-x}{\mu}} \quad (6)$$

Using this mathematical distribution, one can generate reasonable values for transmission line per unit reactance in a given synthetic grid model. Note that, the distribution of per unit reactance for transmission lines is very different from TLS distribution for those of transformers. This is bacuse of per unit conversion for transforemrs and implies that in order to have more stablized range for transmission lines reactance, it is better to study their actual distributed reactance (Ω/km). This will be presented in our next comprehensive study.

### B. Transmission line capacity distribution

Transmission line capacity is a critical parameter in various analysis such as optimal power flow (OPF) analysis, contingency analysis, and power grid expansion planning. Therefore, here we studied the distribution of line capacity for different voltage levels to identify a useful guideline and range for actual capacities in the real grids. Fig. 7 shows the empirical PDF of transmission line capacity and the approximated normal distribution with best estimated parameters based on $D_{KL}$ for three different voltage levels. Note that, unlike transformers the distribution of MVA rating for transmission lines is approximated with normal distribution with higher mean values for each voltage level.

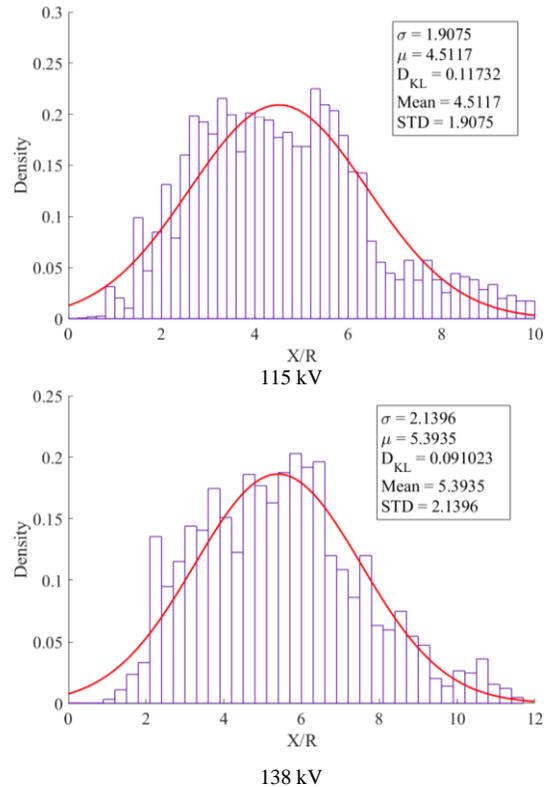

Figure 7. Empirircal PDF and Normal-fit of line capacity for 115, 138, and 230 kV transmission lines.

### C. Transmission line X/R ratio distribution

The third important parameter of transmission lines is the reactance to resistance ratio. Using such ratio, one can estimate the value of per unit resistance given the range and distribution of per unit reactance of the line. These two parameters form the real and imaginary parts of transformer impedance that is necessary for power flow analysis in synthetic power networks. Fig. 8 shows the empirical PDF of transmission line X/R ratio and the approximated distribution with best estimated parameters based on $D_{KL}$ measure for three different voltage levels. It is found that normal distribution is the best fit for this parameter based on the empirical PDF derived from actual data from two power grids. As shown in Fig. 8, the X/R ratio of transmission lines for each voltage level is smaller than that of transformers. Also, note that for both transformers and transmission lines, this ratio grows as the voltage level increases.

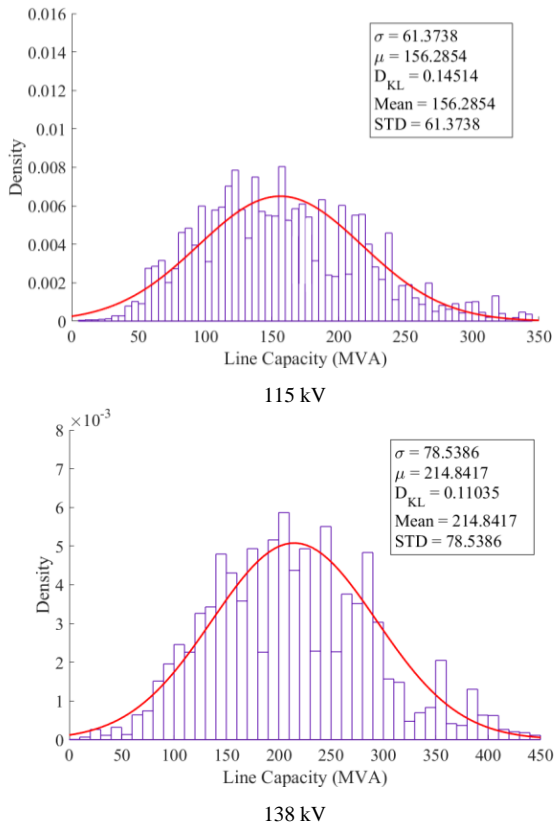

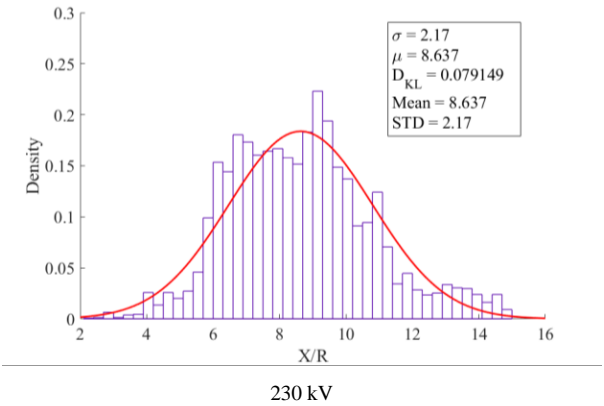

230 kV

Figure 8.  Empirircal PDF and Normal-fit of line X/R ratio for 115, 138, and 230 kV transmission lines.

IV. CONCLUSION AND FUTURE WORKS

Statistical analysis on transformers and transmission lines electrical parameters such as per unit reactance, MVA rating, and X/R ratio is performed in this study to provide both validation metrics and guidelines for generating synthetic grid models. A large sample of real data on transformers and transmission lines from two real-world power systems is used to obtain statistics for the electrical parameters. First, a comparison made between per unit reactance calculated using system common base and values calculated using transformer power base to decide which metric provides more stabilized range for per unit reactance of transformers. It is found that using per unit reactance calculated based on transformer own MVA rating will give us a consistently stabilized range for per unit X over different voltage levels. Next, using Kullback-Leibler divergence, we tried to fit approximate distribution functions on empirical PDFs for electrical parameters of branches. It is found that transmission lines exhibit different statistical properties than transformers. The distribution of X/R ratio for transmission lines is approximated with normal distribution as opposed to the GEV distribution of this parameter for transformers. Also, this ratio is larger for transformers compared to transmission lines. It is also found that transformers/transmission lines of the higher voltage levels tend to have higher power ratings and X/R ratios. Our analyses provide a list of well-defined rules for validation purpose in synthetic grid models. In addition, obtained fit distributions can be used to configure electrical parameters of transmission network in synthetic grid modeling. For future work, a comprehensive study will cover more voltage levels in statistics derivations. We will also study the interdependence of different electrical parameters of transmission network on voltage level.